\documentclass[10pt,twocolumn,letterpaper]{article}

\usepackage{cvpr}              

\usepackage{pifont}
\newcommand{\cmark}{\ding{51}}
\newcommand{\xmark}{\ding{55}}

\definecolor{cvprblue}{rgb}{0.21,0.49,0.74}
\usepackage[pagebackref,breaklinks,colorlinks,allcolors=cvprblue]{hyperref}

\def\paperID{MPI-002} 
\def\confName{BMVC}
\def\confYear{2025}

\title{Modelling the Interplay of Eye-Tracking Temporal Dynamics and Personality for Emotion Detection in Face-to-Face Settings}

\author{
Meisam J.~Seikavandi\\
brAIn Lab, IT University of Copenhagen
\and
Jostein Fimland\\
brAIn Lab, IT University of Copenhagen
\and
Fabricio Batista Narcizo\\
GN Advanced Science
\and
Maria Barrett\\
IT University of Copenhagen
\and
Ted Vucurevich\\
GN Advanced Science
\and
Jesper Bünsow Boldt\\
GN Advanced Science
\and
Andrew Burke Dittberner\\
GN Advanced Science
\and
Paolo Burelli\\
brAIn Lab, IT University of Copenhagen\\
}

\begin{document}
\maketitle

\begin{abstract}
Accurate recognition of human emotions is critical for adaptive human–computer interaction, yet remains challenging in dynamic, conversation-like settings. This work presents a personality-aware multimodal framework that integrates \textbf{eye-tracking sequences}, \textbf{Big Five personality traits}, and \textbf{contextual stimulus cues} to predict both \textit{perceived} and \textit{felt} emotions. Seventy-three participants viewed speech-containing clips from the CREMA-D dataset while providing eye-tracking signals, personality assessments, and emotion ratings. Our neural models captured temporal gaze dynamics and fused them with trait and stimulus information, yielding consistent gains over SVM and literature baselines. Results show that (i) stimulus cues strongly enhance perceived-emotion predictions (macro F1 up to \textbf{0.77}), while (ii) personality traits provide the largest improvements for \textit{felt} emotion recognition (macro F1 up to \textbf{0.58}). These findings highlight the benefit of combining physiological, trait-level, and contextual information to address the inherent subjectivity of emotion. By distinguishing between perceived and felt responses, our approach advances multimodal affective computing and points toward more personalized and ecologically valid emotion-aware systems.
\end{abstract}

\section{Introduction}

Emotion recognition is a central challenge in affective computing, with wide-ranging applications in areas such as adaptive human–computer interaction, virtual agents, education, and teleconferencing. Accurate recognition enables systems to interpret and respond to human states in a more personalized and context-sensitive manner. Yet, despite decades of progress, most computational models still rely on simplified representations—for example, static snapshots of facial expressions or coarse emotion labels—that fall short of capturing the richness of real human interactions. Genuine emotion perception unfolds dynamically, is shaped by subtle attentional cues, and is further modulated by stable individual traits such as personality~\cite{hughes2020personality,kaspar2012emotions,rauthmann2012eyes}.

At the theoretical level, these complexities lie at the intersection of two perspectives. \emph{Basic Emotion Theory} (BET)~\cite{Ekman1992} views emotions as discrete, biologically hard-wired categories, while constructionist accounts such as the \emph{Theory of Constructed Emotion} (TCE)~\cite{barrett2017theory} emphasize that emotions emerge from interactions between core affect and conceptual knowledge. A layered view suggests that both levels matter: an observer can recognize expressed cues in another person (BET), yet their own internal affective experience may be constructed differently (TCE).

This distinction is particularly salient in what we term the \emph{listener’s perspective}. In everyday scenarios—watching a speaker in a meeting, following a video call, or interacting with a virtual agent—people interpret a talking face without necessarily engaging in full turn-taking dialogue. Such contexts approximate a dialogue-like setting while retaining one-sided attentional dynamics. Our study focuses on this \emph{listener scenario}, where participants watched speech-containing video clips from the CREMA-D dataset~\cite{cao2014crema}, a setup that allows us to probe how perceived and felt emotions may converge or diverge.

Key challenges include neglected temporal dynamics, under-modeled individual differences, and divergence between perceived and felt emotions~\cite{barrett2017theory}.

To systematically address these issues, we adopt a framework distinguishing \textit{Expressed Emotions} ($E_e$), \textit{Perceived Emotions} ($E_p$), and \textit{Felt Emotions} ($E_f$). Encountering emotional stimuli involves recognizing $E_e$, forming a subjective perception $E_p$, and potentially experiencing a distinct internal state $E_f$. Modeling both $E_p$ and $E_f$ in parallel is therefore essential for realistic emotion recognition.

In this paper, we introduce a multimodal approach that integrates \textbf{eye-tracking data}, \textbf{temporal modeling}, and \textbf{personality traits} to predict $E_p$ and $E_f$ in a dynamic, speech-based setting. Seventy-three \emph{non-actor participants} contributed (1) detailed eye movements (fixations, pupil size), (2) self-reported Big Five personality profiles~\cite{john1991big,fossati2011big}, and (3) perceived and felt emotional responses on each trial. Our contributions are threefold:
\begin{enumerate}
    \item \textbf{Integration of multimodal data:} Demonstrating that combining eye-tracking signals, personality traits, and temporal dynamics significantly improves emotion recognition in talking-face scenarios.
    \item \textbf{Insights into personality’s role:} Providing empirical evidence that personality traits modulate both how participants \emph{perceive} others’ emotions and how they \emph{feel} in response.
    \item \textbf{Advancement in modeling:} Proposing a multimodal neural architecture that achieves high predictive performance for both perceived and felt emotions, with direct relevance for adaptive, user-centered affective computing.
\end{enumerate}

By jointly modeling dynamic cues, stable personality traits, and the divergence between perceived and felt states, our work moves beyond static or purely acted benchmarks. Although the study does not yet capture fully interactive dialogues, it reflects a critical real-world interaction pattern: emotional decoding by the listener. This perspective, we argue, is essential for developing next-generation emotion-recognition systems that respect individual differences and capture the subtle, layered nature of human emotions.

\section{Background}
A wide range of studies has explored different modalities and methodologies, often achieving impressive accuracy levels. However, many of these works simplify emotion detection by focusing on limited arousal and valence scores or by relying on highly controlled datasets. In practice, real-world emotion perception is shaped by multiple, often interdependent factors such as personality traits, gaze patterns, and contextual cues. Consequently, several important challenges remain insufficiently addressed, especially regarding the use of non-actor participants, individual differences, and temporal dynamics. 
These unresolved issues echo long-standing theoretical debates, e.g.\ whether emotions are discrete biological kinds or context-dependent constructions, highlighting the need for frameworks that link physiology, attention, and conceptual labeling in a single pipeline.

\subsection{Emotion Models}
Two primary frameworks characterize how emotions are commonly modeled.  
\textbf{Discrete models} rooted in \emph{Basic Emotion Theory} (BET)~\cite{Ekman1992} group emotions into basic categories such as anger, disgust, fear, joy, sadness, and surprise. These labels are intuitive and map neatly onto facial action patterns, but they often struggle to capture subtle or mixed states in realistic settings~\cite{siegert2011appropriate}.  
\textbf{Dimensional models}, by contrast, represent affect along continuous axes such as arousal and valence~\cite{mehrabian1996pleasure,Russell1980}, occasionally adding dominance as a third dimension. They allow nuanced representations, yet many engineering studies discretize them into low/medium/high bins, obscuring fine-grained shifts.

\textbf{Trade-offs and hybrid views.} Discrete categories can over-simplify overlapping expressions, whereas purely dimensional schemes cannot easily separate qualitatively distinct emotions that share similar core affect (e.g.\ anger vs.\ fear). Recent evidence shows that subjective reports cluster into at least 27 categories connected by smooth gradients~\cite{cowen2017self}, motivating \emph{hybrid} pipelines that first estimate core affect and then map it onto discrete concepts, an approach compatible with \emph{Theory of Constructed Emotion} (TCE)~\cite{barrett2017theory}.  
Following Van Heijst et al.~\cite{van2025basic}, we view BET and TCE as complementary layers: BET explains why evolution endowed us with affect programs; TCE explains how any given episode is constructed from core affect plus conceptual knowledge. This layered stance underpins our decision to model both continuous ($E_f$) and categorical ($E_p$) labels in later sections.

\subsection{Multimodal Emotion Recognition Approaches}
Leveraging multiple modalities facial expressions, vocal prosody, and physiological signals has yielded robust gains. For example, Kollias et al.~\cite{kollias2022} used multi-task audio-visual learning to detect valence, arousal, expressions, and action units. Reviews by Li et al.~\cite{li2024} and Zhang et al.~\cite{zhang2023} report strong scores in controlled labs.  
Yet most pipelines conflate low-level arousal cues with high-level categorical labels, ignoring the layered distinction between core affect and conceptual emotion. Moreover, many still rely on \emph{actors} or compress ratings to binaries, limiting ecological validity.

Practical constraints persist: collecting EEG or GSR requires specialized hardware, so researchers increasingly explore more accessible channels such as eye tracking or basic speech while still capturing real-world complexity.

\subsection{Eye-Tracking-Based Emotion Recognition}
Among visual modalities, \textbf{eye tracking} uniquely captures both attentional focus (fixations) and arousal (pupil dilation)~\cite{AMultiMohammadi2022}. Eye movements correlate with emotional states~\cite{seikavandi2025modeling} and reveal which facial regions observers deem salient~\cite{skaramagkas_review_2023}. Lu et al.~\cite{lu2015combining} combined eye tracking with EEG to boost accuracy.  
Attachment style and personality dimensions modulate pupillary responses e.g.\ avoidant individuals show blunted dilation to happy faces, underscoring the value of trait-aware models~\cite{vacaru2025attachment}.  
Nevertheless, lighting, eyewear, or calibration drift can degrade data quality.

\subsection{Gaze Strategies}
Eye-gaze strategies how observers allocate fixations across a face offer insight into emotion-specific cues. Patterns vary by age~\cite{chaby2017gaze} and gender~\cite{coutrot2016face}. Different regions (eyes vs.\ mouth) carry diagnostic weight for particular emotions~\cite{schurgin2014eye}, and gaze direction modulates perception~\cite{liang2021emotional}. Systems focusing solely on the eyes may thus miss critical mouth cues, and vice versa.

\subsection{Stimuli and Participant Considerations}
A recurring critique is reliance on \emph{actors} displaying prototypical expressions~\cite{park2024,chen2024}. Such datasets inflate accuracy yet transfer poorly to spontaneous contexts. Recruiting \textbf{non-actors} and using more naturalistic stimuli improves ecological validity but increases variability. Our “talking face” paradigm with non-actor participants aims to balance realism and control.

\subsection{Temporal Dynamics in Emotion Recognition}
Most pipelines still treat emotions as static snapshots, ignoring their evolution. Wang et al.~\cite{wang2023} emphasize the need for sequence modeling to capture rapid affective transitions; without temporal context, fleeting cues may be misinterpreted.

\subsection{Personality and Emotion Recognition}
Personality, typically the Big Five shapes both expression and perception~\cite{kehoe2012personality,zautra2005dynamic}. Neurotic individuals fixate on negative content; extraverts seek positive cues~\cite{costa1980influence}. Eye-tracking studies now infer personality traits themselves~\cite{ait2018gaze,rauthmann2012eyes,chen_eye-tracking-based_2023}, paving the way for adaptive systems. These advances raise privacy concerns and demand careful trait inference.

\subsection{Personality-Inspired Eye-Tracking-Based Emotion Recognition}
Incorporating personality scores can improve gaze-based emotion recognition. High-neuroticism observers linger on negative features~\cite{chen2023emotion}; extraverts scan positive cues~\cite{haas2008stop}. Real-world validations remain scarce, and challenges include maintaining calibration and safeguarding privacy. Nevertheless, personality-aware pipelines represent a stride toward user-centric, empathetic computing.

In sum, emotion-recognition pipelines still over-rely on acted stimuli, neglect individual differences, and ignore layered temporal dynamics—gaps our study addresses by combining naturalistic “talking-face” stimuli with eye-tracking and personality traits in a BET–TCE framework.

\section{Dataset Collection and Preprocessing}

\subsection{Participants}
We recruited 73 participants (52 males, 21 females; mean age $27.4 \pm 6$ years). All participants reported normal or corrected-to-normal vision and no neurological disorders. Participants came from diverse educational backgrounds (see Table~\ref{tab:demographics}). The participants agreed and signed the informed consent following the university's ethical guidelines. Although this sample provides educational diversity, we note a moderate gender imbalance and a relatively young average age, which may limit broader generalizability.

\begin{table}[h]
\caption{Participant Demographics}
\centering
\begin{tabular}{l c}
\hline
\textbf{Characteristic} & \textbf{Value} \\
\hline
Gender (Male/Female) & 72\% / 28\% \\
Average Age (years) & 27.4 $\pm$ 6 \\
Glasses/Contact Lens Wearers & 33\% \\
Education Level: & \\
\quad College & 28\% \\
\quad Bachelor's & 43\% \\
\quad Master's & 19\% \\
\quad Ph.D. & 10\% \\
\hline
\end{tabular}
\label{tab:demographics}
\end{table}

\subsection{Experimental Design and Procedure}
\begin{figure}[htb]
    \centering
    \includegraphics[width=0.40\textwidth]{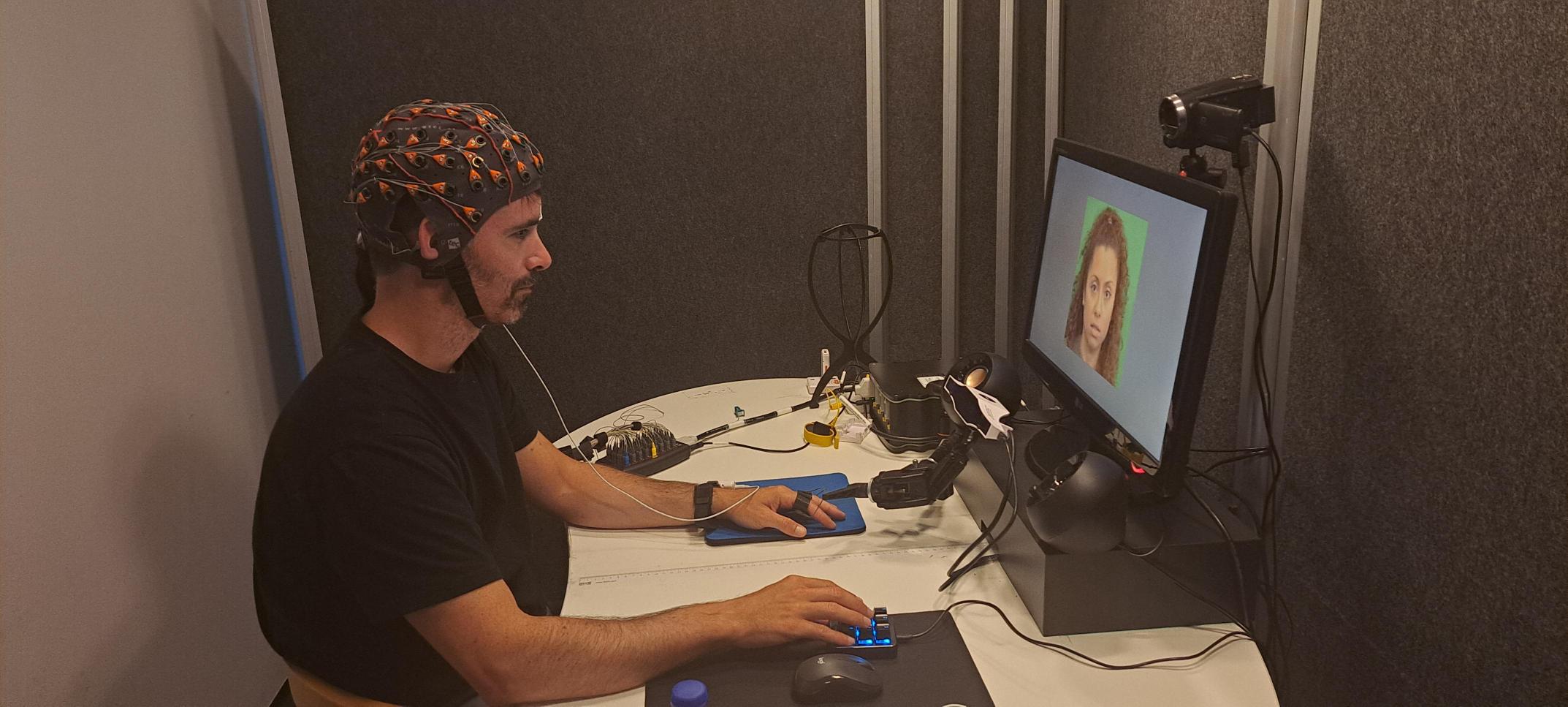}
    \caption{The experimental setup of a participant seated in front of the monitor with sensors attached.
       }
    \label{fig:experiment_setup}
\end{figure}

We simulated the listening aspect of a conversational setting where participants engaged with dynamic, emotionally expressive stimuli. Each participant completed 88 trials (4 practice and 84 main) in random order. Stimuli were 84 video clips from the CREMA-D dataset~\cite{cao2014crema}, featuring 91 actors (48 male, 43 female) aged 20--74, each portraying one of six basic emotions (Anger, Disgust, Fear, Happy, Neutral, Sad) at varying intensities. The selected clips balanced emotions and actor demographics to enhance expressiveness and generalizability.

To approximate face-to-face interaction, a short written scenario was displayed before each video, prompting participants to imagine conversing with the individual shown. This contextual priming aimed to increase engagement and emotional alignment, despite the lack of true turn-taking. The use of short textual prompts was intended not only to simulate real conversational framing but also to standardize participants’ cognitive approach, ensuring consistent engagement across trials rather than replicating spontaneous dialogues.

Eye-tracking data were recorded using a GP3 HD eye tracker at 150 Hz. The eye tracker was calibrated for each participant with a standard 9-point procedure. We synchronized data collection with stimulus presentation via the Lab Streaming Layer (LSL) to ensure precise alignment between eye-tracking data and stimulus onset.

Figure~\ref{fig:experiment_setup} illustrates the experimental setup, with the participant seated in front of the monitor and wearing sensors.

Before starting the trials, participants completed the BFI-44 questionnaire~\cite{fossati2011big} to assess openness, conscientiousness, extraversion, agreeableness, and neuroticism. After each video, participants rated their \emph{perceived} and \emph{felt} emotions on 9-point Likert scales for valence (1 = very negative, 9 = very positive) and arousal (1 = very calm, 9 = very excited). We chose a 9-point scale for its higher resolution, capturing more subtle affective nuances~\cite{lang2019affective,benitez2022likert} compared to smaller scales. These self-reported ratings form the ground truth labels for our emotion recognition models.

\begin{figure}[htb]
    \centering
    \includegraphics[width=0.25\textwidth]{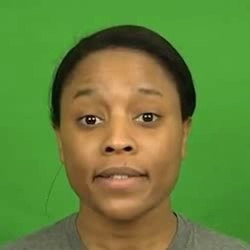}
    \caption{A frame from the CREMA-D dataset~\cite{cao2014crema} used in the experiment.}
    \label{fig:example_video_stimuli}
\end{figure}

\begin{figure}[htb]
    \centering
    \includegraphics[width=0.4\textwidth]{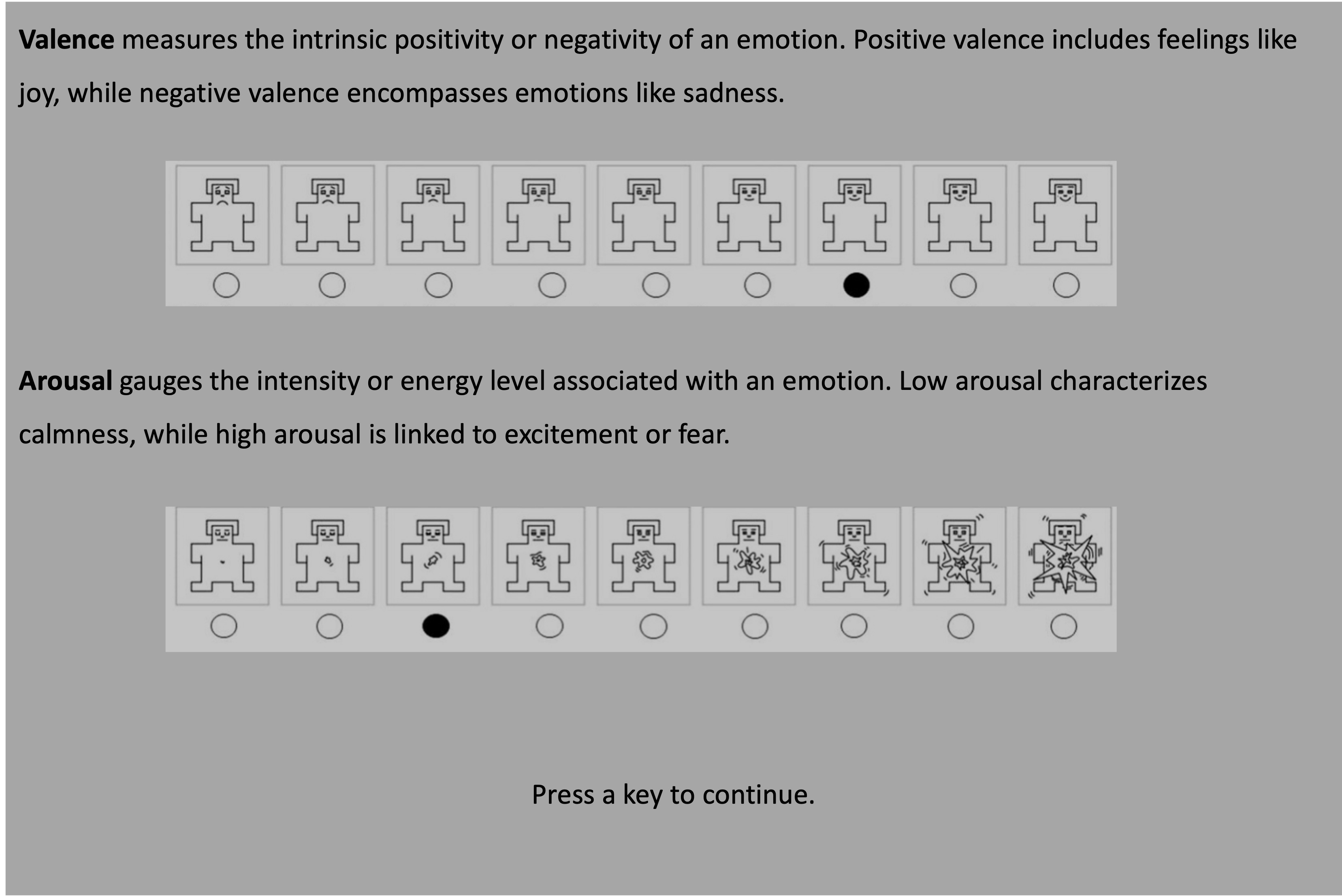}
    \caption{The 9-point scales used to rate emotional arousal and valence.
    }
    \label{fig:likert_scales}
\end{figure}

\subsection{Data Preprocessing and Feature Extraction}
Preprocessing included (i) quality filtering (removing blinks/tracking loss), (ii) normalization of gaze coordinates, and (iii) baseline correction of pupil size relative to neutral trials.

\begin{figure}[t]
    \centering
    \includegraphics[width=0.3\textwidth]{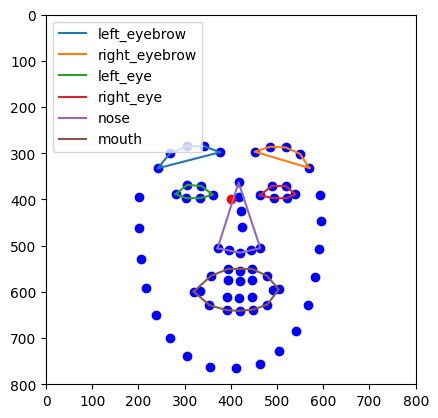}
    \caption{Facial landmarks extracted by OpenFace~\cite{amos2016openface} partitioned into multiple ROIs.}
    \label{fig:segmentation}
\end{figure}

For each trial, we extracted:
\begin{itemize}
    \item \textbf{Fixation Metrics:} Mean, median, and variance of fixation duration/dispersion.
    \item \textbf{Pupil Metrics:} Corrected mean, min, max, and variance of pupil size.
    \item \textbf{Saccadic Metrics:} Amplitude, duration, peak velocity, and acceleration, representing scanning behavior.
    \item \textbf{Gaze Regions:} Using OpenFace~\cite{amos2016openface} and convex hull segmentation~\cite{adipranata2009fast}, we labeled each gaze point by region (eyes, eyebrows, nose, mouth, or outside). The proportion of fixations in each region provides insight into attentional distribution.
    \item \textbf{Environmental Variables:} Ambient light (lux), room temperature, and stimulus brightness levels to control for extraneous influences on pupil dilation.
    \item \textbf{Personality Traits:} BFI-44 scores scaled to [0,1].
\end{itemize}

These features span multiple modalities: dynamic eye-tracking, static personality and environment, and categorical stimulus labels, and are summarized in Table~\ref{tab:feature_inventory}.

\begin{table}[t]
\caption{Feature Inventory}
\vspace{-5pt}
\resizebox{\linewidth}{!}{
\label{tab:feature_inventory}
\centering
\begin{tabular}{lccc}
\hline
Modality      & Variables                   & Dim.    & Temporal \\
\hline
Eye-tracking  & FixDur, PupilMean, …        & 15×12   & \cmark         \\
Personality   & Big-Five (O, C, E, A, N)     & 5       & \xmark         \\
Stimulus emo  & One-hot (6)                 & 6       & \xmark         \\
Environment   & Lux, Temp          & 2       & \xmark         \\
\hline
\end{tabular}}
\end{table}

\subsection{Features for Modeling}
Because fixations and saccades vary in frequency and timing, we standardized the temporal dimension via \textbf{interpolation} into 15 equally spaced time steps per trial. This uniform representation accommodates 2–4 second videos and facilitates sequential modeling with architectures such as Long Short-Term Memory (LSTM) networks~\cite{graves2005bidirectional}.  
Interpolating to equal-length sequences allows us to treat each trial as a short “emotion episode” in the layered-affect sense, i.e., a window where core-affect fluctuations (pupil, arousal) can be mapped to conceptual labels.  

Each time step includes gaze-region allocations, pupil size, and saccadic measures, capturing how attention and arousal evolve over time. By integrating these time-series features with static context variables (environment and personality), our models capitalize on both dynamic and trait-level information to boost emotion recognition accuracy.

\section{Machine Learning Modeling}

\subsection{Emotion Labeling and Data Preparation}
We aimed to predict four emotion labels: \textbf{felt valence}, \textbf{perceived valence}, \textbf{felt arousal}, and \textbf{perceived arousal}. Given the imprecision in self-reported data, each label was grouped into three classes: low/negative (1--3), medium/neutral (4--6), and high/positive (7--9). We divided the dataset into training (64\%), validation (16\%), and testing (20\%) subsets using stratified splits to maintain class distribution.
These splits were not strictly subject-independent. To reduce the risk of personality vectors being memorized across folds, we injected small random Gaussian noise into personality scores during training for each trial, which acted as a regularization strategy.

While this binning approach improves model stability, it may mask finer affective distinctions, so future research could explore regression-based or ordinal classification. Following prior studies~\cite{fiorini2024eeg,heffner2022probabilistic,garg2022decoding}, binning continuous ratings also mitigates subjective variability in self-reported emotions. Nonetheless, some granularity is inevitably lost.

\subsection{Feature Engineering and Preprocessing}
\subsubsection*{Normalization and Scaling}
To ensure fair feature contribution and reduce bias, we applied consistent preprocessing. Personality trait scores (0--50) were scaled by dividing by 50~\cite{Scikit-learn}. Highly skewed features, like saccade amplitude/duration, were transformed with \texttt{MinMaxScaler}. Other continuous features (pupil sizes, environment variables) were standardized using \texttt{StandardScaler}, subtracting the training-set mean and dividing by its standard deviation to avoid leakage.

\subsubsection*{One-Hot Encoding}
We encoded stimulus emotion (happy, sad, neutral, angry, disgust, fear) as a 6-dimensional one-hot vector, allowing the model to distinguish each emotion category independently.

\subsection{Neural Network Architecture}
\begin{figure}[htb!]
\centering
\includegraphics[width=0.45\textwidth]{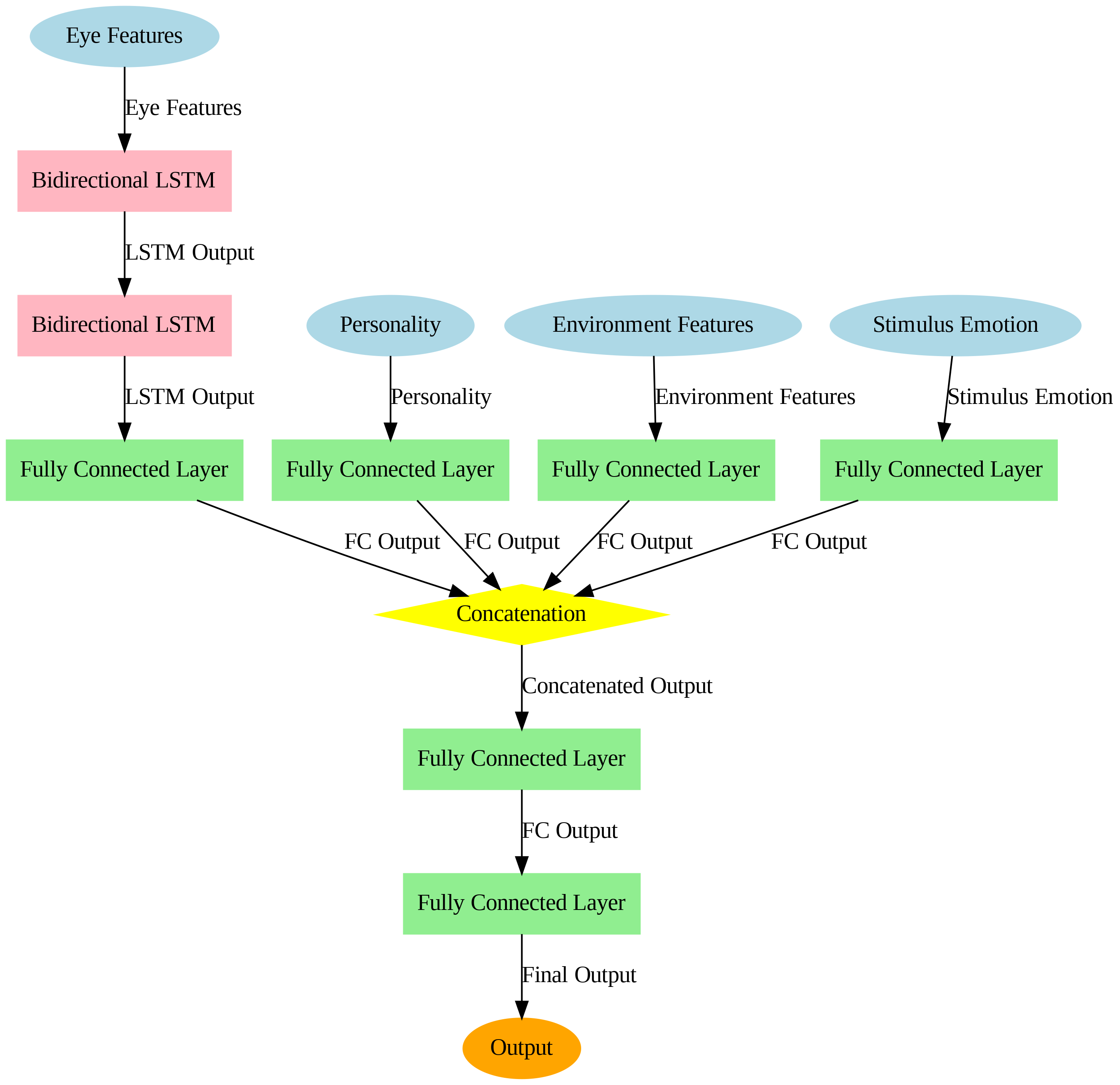}
\caption{Neural network architecture integrating eye-tracking data, environmental variables, personality traits, and stimulus emotion.}
\label{fig:model_architecture}
\end{figure}

Our neural network (NN) integrates multiple input streams (Figure~\ref{fig:model_architecture}). Each stream undergoes separate preprocessing before feature fusion. This design ensures that temporal features (e.g., eye-tracking data) and static features (e.g., personality traits, stimulus emotion) receive tailored treatment for their respective roles in predicting emotional states.

\noindent \textbf{Eye-Tracking Data}: Processed through LSTM layers to capture temporal dependencies.\\
\textbf{Personality Traits}: Processed through fully connected layers.\\
\textbf{Stimulus Emotion}: One-hot encoded and passed through fully connected layers.\\
\textbf{Environmental Variables}: Processed through fully connected layers.

\noindent To mitigate overfitting on personality or environmental variables, we injected small Gaussian noise into these inputs during training as a form of data augmentation. This noise helps the model generalize across participants and conditions.

\subsection{Classification Approach}
We framed emotion prediction as a three-class classification task, applying softmax activation for class probabilities. We used categorical cross-entropy loss with inversely proportional class weights to handle imbalance.

\subsection{Model Training and Evaluation}
We performed manual and grid search hyperparameter tuning for our neural networks. The search space covered learning rates $\{10^{-3}, 10^{-4}, 10^{-5}\}$, dropout rates $\{0.2, 0.3, 0.5\}$, and weight decay values. We selected the best configurations based on macro F1 scores on the validation set. Training employed early stopping, halting if validation did not improve within 10 epochs.

Hyperparameters like learning rate, dropout, and weight decay were chosen per validation performance. We used the F1-score for evaluation due to dataset imbalance, as it considers both precision and recall.

We compared our NN models with support vector machines (SVMs) as baselines. 
The SVMs used stimulus emotion alone or combined with personality data. 
We chose SVM because it is simple, effective for non-temporal data, and reveals the benefit of adding sequential modeling in the NN. 
\paragraph{Baseline limitation.}
Our SVM baseline used aggregated features without temporal gaze dynamics.
A stricter comparison with identical feature sets would further strengthen causal attribution of performance gains, but was beyond the scope of this paper.

Although this comparison demonstrates the added value of temporal modeling, a full ablation—training both NN and SVM on identical feature sets—was beyond the scope of this paper but remains a clear next step for strengthening causal attribution of performance gains.

\subsection{Results}
Table~\ref{tab:combined_model_performance} shows the F1-scores for different models and emotion labels, with the best results for each label \textbf{bolded}.

\begin{table*}
	\centering
    \caption{Model performance (F1-scores) and hyperparameters (Learning Rate and Dropout) for different input features.}
	\begin{tabular}{lcccccc}
    	\hline
    	& \textbf{Low} & \textbf{Medium} & \textbf{High} & \textbf{Macro F1} & \textbf{Learning Rate} & \textbf{Dropout} \\
    	\hline
    	\textbf{NN with Eye-Tracking Data (No Env)} & & & & & & \\
    	Perceived Arousal & 0.32 & 0.54 & 0.17 & 0.34 & 0.0002 & 0.3 \\
    	Perceived Valence & 0.35 & 0.22 & 0.28 & 0.28 & 0.0002 & 0.3 \\
    	Felt Arousal & 0.45 & 0.37 & 0.07 & 0.30 & 0.0003 & 0.2 \\
    	Felt Valence & 0.29 & 0.49 & 0.24 & 0.34 & 0.0002 & 0.3 \\
    	\hline
    	\textbf{NN with Eye-Tracking Data} & & & & & & \\
    	Perceived Arousal & 0.18 & 0.57 & 0.24 & 0.33 & 0.00035 & 0.3 \\
    	Perceived Valence & 0.58 & 0.17 & 0.29 & 0.34 & 0.00035 & 0.3 \\
    	Felt Arousal & 0.45 & 0.41 & 0.23 & 0.36 & 0.0003 & 0.2 \\
    	Felt Valence & 0.32 & 0.46 & 0.25 & 0.34 & 0.0003 & 0.2 \\
    	\hline
    	\textbf{NN with Eye-Tracking + Personality} & & & & & & \\
    	Perceived Arousal & 0.46 & 0.49 & 0.40 & 0.45 & 0.0003 & 0.2 \\
    	Perceived Valence & 0.57 & 0.33 & 0.29 & 0.40 & 0.0002 & 0.2 \\
    	Felt Arousal & 0.58 & 0.58 & 0.40 & \textbf{0.52} & 0.0002 & 0.2 \\
    	Felt Valence & 0.38 & 0.57 & 0.28 & 0.41 & 0.0002 & 0.2 \\
    	\hline
    	\textbf{NN with Eye-Tracking + Stimuli Emotion} & & & & & & \\
    	Perceived Arousal & 0.56 & 0.33 & 0.58 & 0.49 & 0.0002 & 0.3 \\
    	Perceived Valence & 0.77 & 0.56 & 0.91 & 0.75 & 0.0002 & 0.3 \\
    	Felt Arousal & 0.47 & 0.45 & 0.29 & 0.40 & 0.0002 & 0.2 \\
    	Felt Valence & 0.50 & 0.51 & 0.54 & \textbf{0.52} & 0.0003 & 0.3 \\
    	\hline
    	\textbf{NN with Eye-Tracking + Personality + Stimuli} & & & & & & \\
    	Perceived Arousal & 0.63 & 0.48 & 0.65 & \textbf{0.59} & 0.0007 & 0.3 \\
    	Perceived Valence & 0.77 & 0.63 & 0.90 & \textbf{0.77} & 0.0007 & 0.3 \\
    	Felt Arousal & 0.61 & 0.53 & 0.48 & \textbf{0.54} & 0.0004 & 0.3 \\
    	Felt Valence & 0.53 & 0.62 & 0.60 & \textbf{0.58} & 0.0007 & 0.3 \\
    	\hline
    	\textbf{SVM with Stimuli Emotion} & & & & & & \\
    	Perceived Arousal & 0.57 & 0.26 & 0.61 & 0.48 & N/A & N/A \\
    	Perceived Valence & 0.76 & 0.52 & 0.92 & 0.73 & N/A & N/A \\
    	Felt Arousal & 0.48 & 0.28 & 0.32 & 0.36 & N/A & N/A \\
    	Felt Valence & 0.50 & 0.36 & 0.59 & 0.48 & N/A & N/A \\
    	\hline
	\end{tabular}
\label{tab:combined_model_performance}
\end{table*}

Integrating personality traits, temporal eye-tracking data, and stimulus emotion notably boosted performance, especially for \textbf{felt emotions}. This finding suggests that subjective felt experiences profit most from incorporating high-level personality data. The SVM baselines performed well on perceived emotions, possibly reflecting direct stimulus influence, but they could not model sequential dependencies.

\section{Discussion}
We combined eye-tracking data, personality traits, and stimulus-emotion labels to enhance emotion recognition in short speech-containing clips. Although not a fully interactive setup, emphasizing the listener’s perspective under controlled conditions allowed us to isolate key predictors of perceived and felt emotions.

\subsection{Complexity and Agreement}
Emotions emerge from the interplay of stimuli, individual traits, and context, making them challenging to model. Table~\ref{tab:user_agreement} shows user agreement ranging from 56.0\% (felt arousal) to 77.7\% (perceived valence). Personalized calibration could help address subjective variability. Moreover, as gaze patterns vary by age and gender, our demographic imbalance may introduce bias and limit generalizability.

\begin{table}[b]
\centering
\caption{User Agreement on Emotion Labels (\%)}
\begin{tabular}{lc}
\hline
\textbf{Emotion Label} & \textbf{Agreement (\%)} \\
\hline
Felt Arousal & 56.0 \\
Felt Valence & 65.9 \\
Perceived Arousal & 60.6 \\
Perceived Valence & 77.7 \\
\hline
\end{tabular}
\label{tab:user_agreement}
\vspace{-6pt}
\end{table}

\subsection{Model Performance and Stimulus Emotion}
The inclusion of stimulus emotion as an input is motivated by its role as a contextual prior: it represents the actor’s intended expression ($E_e$), which observers can perceive ($E_p$) and relate to their own felt states ($E_f$). 
This is not equivalent to ground truth but serves as a contextual feature that participants explicitly rated against. 
We deliberately did not use raw audiovisual features from the video, as our goal was to focus on observer-centric signals (gaze, personality) rather than re-training an audio–visual recognition system on acted data already well-studied in prior work.

Our best model attained a macro F1 of \textbf{0.77} for perceived valence (Table~\ref{tab:combined_model_performance}). Stimulus emotion greatly aided perceived-emotion prediction; an SVM with only stimulus emotion was already strong. However, the NN outperformed it on felt emotions by integrating physiological cues from eye tracking.  
The fact that perceived valence is easier mirrors findings that observers rely on learned emotion concepts (BET-like), whereas felt states reflect constructionist variability~\cite{Ekman1992,barrett2017theory}.

\subsection{Multimodality and Individual Differences}
Combining stimulus emotion, personality, and eye-tracking gave the highest macro F1 scores (0.77 for perceived valence, 0.58 for felt valence). Personality clarified individual tendencies, eye tracking offered real-time physiological measures~\cite{tarnowski2020eye}, and stimulus emotion contextualized perception.  
This pattern aligns with a layered framework of affect: physiological arousal signals (e.g., pupil dilation) guide attentional deployment (gaze patterns), which in turn feed into conceptual emotion labeling in the observer’s mind~\cite{van2025basic}.

\subsection{Comparison and Future Directions}
While our NN surpassed the SVM baseline for felt emotions—largely due to sequential information—we intentionally kept the LSTM architecture compact to reduce overfitting risk given the dataset size and short sequence lengths.
Future work could evaluate transformer-based multimodal architectures or cross-modal attention mechanisms, which may better exploit multimodal dependencies.
This decision was intentional: the dataset size and the relatively short temporal sequences favored compact architectures with fewer parameters, reducing the risk of overfitting. 
Future work could evaluate transformer-based multimodal architectures or cross-modal attention mechanisms, which may exploit richer interdependencies.
A thorough ablation would align features for both models. Additional future work includes improving sample representativeness, investigating true two-way interactions, and refining interpretability via attention weighting or feature ablation. Real-world applications (e.g., telehealth or adaptive tutoring) must also respect data privacy and informed consent. The reliance on the acted CREMA-D dataset limits ecological validity, and future research should pursue more spontaneous, diverse stimuli.

\section{Conclusion}
Grounded in a layered affect framework that bridges Basic Emotion Theory and constructionist accounts, this study shows that integrating \emph{temporal eye-tracking}, \emph{personality traits}, and \emph{stimulus emotion} can substantially improve emotion recognition in short, speech-based clips. By emphasizing the \emph{listener’s perspective}, we demonstrated that personality traits enhanced predictions of felt emotions (e.g., felt arousal rose from 0.36 to 0.52), while stimulus emotion strongly supported perceived-emotion performance (perceived valence from 0.34 to 0.77). These results highlight the value of separating core-affect dynamics from conceptual labeling when modeling both perceived and felt emotional states.

The implications are twofold. First, unifying physiological signals (pupil size, gaze), attentional strategies (fixation patterns), and contextual traits (personality, stimulus cues) provides a richer and more individualized account of how people perceive and experience emotions. Such integration has direct applications in user-centric technologies, including adaptive virtual agents, teleconferencing systems, and mental health tools. Second, the findings support a layered theoretical perspective in which physiological fluctuations, attentional deployment, and conceptual knowledge jointly shape emotion construction.

At the same time, several limitations must be acknowledged. Our reliance on acted portrayals from the CREMA-D dataset constrains ecological validity, and the participant pool was skewed toward younger male participants, which reduces demographic generalizability. Furthermore, discretizing continuous self-reports into three bins stabilized training but sacrificed emotional nuance. Addressing these challenges will require larger and more diverse samples, spontaneous and interactive dialogue data, and models capable of handling continuous or ordinal emotion ratings. Expanding beyond eye tracking to include additional modalities such as prosody or micro-expressions will also strengthen ecological validity.

In outlook, this work suggests that effective emotion recognition systems must explicitly separate—and then reintegrate—core-affect signals, attentional strategies, and conceptual constructs. By doing so, they can move beyond static, actor-driven benchmarks toward interactive, real-time, and ethically responsible applications. We envision that adopting this layered perspective will foster more adaptive and privacy-conscious affective computing solutions that better capture the complexity and subjectivity of human emotions.

\section{Ethical Impact Statement}
This research investigates emotion detection in dialogues by integrating eye-tracking data, temporal dynamics, and personality traits. As the study involves human participants, it was conducted with oversight from an ethical review board. Informed consent was obtained from all participants, clarifying data collection, usage, and analysis. All data were anonymized, and participants were informed of their right to withdraw at any time without consequence.

Potential risks include privacy concerns related to emotion recognition, especially for metrics like pupil size that participants cannot consciously control. Unlike facial or vocal expressions, eye metrics such as pupil dilation have minimal cultural awareness, raising the risk of unintentionally revealing emotional states. We addressed these concerns by anonymizing data, restricting data access to authorized personnel, and clearly explaining data usage to participants.

The anonymization of data, explicit communication of its purpose, and careful ethical handling are key mitigation strategies. Our findings may enable more sensitive, context-aware affective computing applications that respect user privacy while advancing the field of emotion recognition in a safe, ethically responsible manner.

{\small
\bibliographystyle{ieeenat_fullname}
\bibliography{main}

\begin{thebibliography}{45}
\providecommand{\natexlab}[1]{#1}
\providecommand{\url}[1]{\texttt{#1}}
\expandafter\ifx\csname urlstyle\endcsname\relax
  \providecommand{\doi}[1]{doi: #1}\else
  \providecommand{\doi}{doi: \begingroup \urlstyle{rm}\Url}\fi

\bibitem[Adipranata et~al.(2009)Adipranata, Ballangan, Ongkodjojo,
  et~al.]{adipranata2009fast}
Rudy Adipranata, Cherry~G Ballangan, Ronald~P Ongkodjojo, et~al.
\newblock Fast method for multiple human face segmentation in color image.
\newblock \emph{International Journal of Advanced Science and Technology},
  3:\penalty0 19--32, 2009.

\bibitem[Ait~Challal and Grynszpan(2018)]{ait2018gaze}
Tagduda Ait~Challal and Ouriel Grynszpan.
\newblock What gaze tells us about personality.
\newblock In \emph{Proceedings of the 6th International Conference on
  Human-Agent Interaction}, pages 129--137, 2018.

\bibitem[Amos et~al.(2016)Amos, Ludwiczuk, Satyanarayanan,
  et~al.]{amos2016openface}
Brandon Amos, Bartosz Ludwiczuk, Mahadev Satyanarayanan, et~al.
\newblock Openface: A general-purpose face recognition library with mobile
  applications.
\newblock \emph{CMU School of Computer Science}, 6\penalty0 (2):\penalty0 20,
  2016.

\bibitem[Barrett(2017)]{barrett2017theory}
Lisa~Feldman Barrett.
\newblock The theory of constructed emotion: an active inference account of
  interoception and categorization.
\newblock \emph{Social cognitive and affective neuroscience}, 12\penalty0
  (1):\penalty0 1--23, 2017.

\bibitem[Benitez-Quiroz et~al.(2022)Benitez-Quiroz, Wilbur, and
  Martinez]{benitez2022likert}
C.~Felipe Benitez-Quiroz, Ronnie~B. Wilbur, and Aleix~M. Martinez.
\newblock Improving the measurement of emotional responses with fine-grained
  likert scales.
\newblock \emph{Emotion Review}, 14\penalty0 (1):\penalty0 26--36, 2022.

\bibitem[Cao et~al.(2014)Cao, Cooper, Keutmann, Gur, Nenkova, and
  Verma]{cao2014crema}
Houwei Cao, David~G Cooper, Michael~K Keutmann, Ruben~C Gur, Ani Nenkova, and
  Ragini Verma.
\newblock Crema-d: Crowd-sourced emotional multimodal actors dataset.
\newblock \emph{IEEE transactions on affective computing}, 5\penalty0
  (4):\penalty0 377--390, 2014.

\bibitem[Chaby et~al.(2017)Chaby, Hupont, Avril, Luherne-du Boullay, and
  Chetouani]{chaby2017gaze}
Laurence Chaby, Isabelle Hupont, Marie Avril, Viviane Luherne-du Boullay, and
  Mohamed Chetouani.
\newblock Gaze behavior consistency among older and younger adults when looking
  at emotional faces.
\newblock \emph{Frontiers in Psychology}, 8:\penalty0 548, 2017.

\bibitem[Chen et~al.(2023{\natexlab{a}})Chen, Cai, Yan, and
  Berkovsky]{chen_eye-tracking-based_2023}
Li Chen, Wanling Cai, Dongning Yan, and Shlomo Berkovsky.
\newblock Eye-tracking-based personality prediction with recommendation
  interfaces.
\newblock \emph{User Modeling and User-Adapted Interaction}, 33\penalty0
  (1):\penalty0 121--157, 2023{\natexlab{a}}.

\bibitem[Chen et~al.(2023{\natexlab{b}})Chen, Liu, Weng, Huang, Weng, Zeng, Li,
  Zheng, and Chen]{chen2023emotion}
Ling Chen, Xiqin Liu, Xiangrun Weng, Mingzhu Huang, Yuhan Weng, Haoran Zeng,
  Yifan Li, Danna Zheng, and Caiqi Chen.
\newblock The emotion regulation mechanism in neurotic individuals: The
  potential role of mindfulness and cognitive bias.
\newblock \emph{International Journal of Environmental Research and Public
  Health}, 20\penalty0 (2):\penalty0 896, 2023{\natexlab{b}}.

\bibitem[Chen and Liu(2024)]{chen2024}
Rui Chen and Qing Liu.
\newblock esee-d: Emotional state estimation based on eye-tracking dataset.
\newblock \emph{ArXiv Preprint}, arXiv:2403.11590, 2024.

\bibitem[Costa and McCrae(1980)]{costa1980influence}
Paul~T Costa and Robert~R McCrae.
\newblock Influence of extraversion and neuroticism on subjective well-being:
  happy and unhappy people.
\newblock \emph{Journal of personality and social psychology}, 38\penalty0
  (4):\penalty0 668, 1980.

\bibitem[Coutrot et~al.(2016)Coutrot, Binetti, Harrison, Mareschal, and
  Johnston]{coutrot2016face}
Antoine Coutrot, Nicola Binetti, Charlotte Harrison, Isabelle Mareschal, and
  Alan Johnston.
\newblock Face exploration dynamics differentiate men and women.
\newblock \emph{Journal of vision}, 16\penalty0 (14):\penalty0 16--16, 2016.

\bibitem[Cowen and Keltner(2017)]{cowen2017self}
Alan~S Cowen and Dacher Keltner.
\newblock Self-report captures 27 distinct categories of emotion bridged by
  continuous gradients.
\newblock \emph{Proceedings of the national academy of sciences}, 114\penalty0
  (38):\penalty0 E7900--E7909, 2017.

\bibitem[Ekman(1992)]{Ekman1992}
Paul Ekman.
\newblock An argument for basic emotions.
\newblock \emph{Cognition \& Emotion}, 6\penalty0 (3-4):\penalty0 169--200,
  1992.

\bibitem[Fiorini et~al.(2024)Fiorini, Bossi, and Di~Gruttola]{fiorini2024eeg}
Linda Fiorini, Francesco Bossi, and Francesco Di~Gruttola.
\newblock Eeg-based emotional valence and emotion regulation classification: a
  data-centric and explainable approach.
\newblock \emph{Scientific Reports}, 14\penalty0 (1):\penalty0 24046, 2024.

\bibitem[Fossati et~al.(2011)Fossati, Borroni, Marchione, and
  Maffei]{fossati2011big}
Andrea Fossati, Serena Borroni, Donatella Marchione, and Cesare Maffei.
\newblock The big five inventory (bfi).
\newblock \emph{European Journal of Psychological Assessment}, 2011.

\bibitem[Garg et~al.(2022)Garg, Garg, Anand, and Baths]{garg2022decoding}
Nikhil Garg, Rohit Garg, Apoorv Anand, and Veeky Baths.
\newblock Decoding the neural signatures of valence and arousal from portable
  eeg headset.
\newblock \emph{Frontiers in Human Neuroscience}, 16:\penalty0 1051463, 2022.

\bibitem[Graves et~al.(2005)Graves, Fern{\'a}ndez, and
  Schmidhuber]{graves2005bidirectional}
Alex Graves, Santiago Fern{\'a}ndez, and J{\"u}rgen Schmidhuber.
\newblock Bidirectional lstm networks for improved phoneme classification and
  recognition.
\newblock In \emph{International conference on artificial neural networks},
  pages 799--804. Springer, 2005.

\bibitem[Haas et~al.(2008)Haas, Constable, and Canli]{haas2008stop}
Brian~W Haas, R~Todd Constable, and Turhan Canli.
\newblock Stop the sadness: Neuroticism is associated with sustained medial
  prefrontal cortex response to emotional facial expressions.
\newblock \emph{Neuroimage}, 42\penalty0 (1):\penalty0 385--392, 2008.

\bibitem[Heffner and FeldmanHall(2022)]{heffner2022probabilistic}
Joseph Heffner and Oriel FeldmanHall.
\newblock A probabilistic map of emotional experiences during competitive
  social interactions.
\newblock \emph{Nature communications}, 13\penalty0 (1):\penalty0 1718, 2022.

\bibitem[Hughes et~al.(2020)Hughes, Kratsiotis, Niven, and
  Holman]{hughes2020personality}
David~J Hughes, Ioannis~K Kratsiotis, Karen Niven, and David Holman.
\newblock Personality traits and emotion regulation: A targeted review and
  recommendations.
\newblock \emph{Emotion}, 20\penalty0 (1):\penalty0 63, 2020.

\bibitem[John et~al.(1991)John, Donahue, and Kentle]{john1991big}
Oliver~P John, Eileen~M Donahue, and Robert~L Kentle.
\newblock Big five inventory.
\newblock \emph{Journal of personality and social psychology}, 1991.

\bibitem[Kaspar and K{\"o}nig(2012)]{kaspar2012emotions}
Kai Kaspar and Peter K{\"o}nig.
\newblock Emotions and personality traits as high-level factors in visual
  attention: a review.
\newblock \emph{Frontiers in human neuroscience}, 6:\penalty0 321, 2012.

\bibitem[Kehoe et~al.(2012)Kehoe, Toomey, Balsters, and
  Bokde]{kehoe2012personality}
Elizabeth~G Kehoe, John~M Toomey, Joshua~H Balsters, and Arun~LW Bokde.
\newblock Personality modulates the effects of emotional arousal and valence on
  brain activation.
\newblock \emph{Social cognitive and affective neuroscience}, 7\penalty0
  (7):\penalty0 858--870, 2012.

\bibitem[Kollias et~al.(2022)Kollias, Tzirakis, Nicolaou, Papaioannou, Zhao,
  Schuller, and Zafeiriou]{kollias2022}
Dimitrios Kollias, Panagiotis Tzirakis, Mihalis~A Nicolaou, Andreas
  Papaioannou, Guoying Zhao, Björn Schuller, and Stefanos Zafeiriou.
\newblock Abaw: Valence-arousal estimation, expression recognition, action unit
  detection \& multi-task learning challenges.
\newblock In \emph{Proceedings of the IEEE/CVF Conference on Computer Vision
  and Pattern Recognition Workshops (CVPRW)}, pages 10790--10800, 2022.

\bibitem[Lang et~al.(2019)Lang, Bradley, and Cuthbert]{lang2019affective}
Peter~J. Lang, Margaret~M. Bradley, and Bruce~N. Cuthbert.
\newblock Affective norms for english words (anew): Affective ratings of words
  and instructions for use.
\newblock \emph{Behavior Research Methods}, 51\penalty0 (4):\penalty0
  1246--1265, 2019.

\bibitem[Li and Chen(2024)]{li2024}
Xuan Li and Yan Chen.
\newblock Emotion recognition using different sensors, emotion models, methods,
  and datasets: A comprehensive review.
\newblock \emph{Frontiers in Neuroergonomics}, 5\penalty0 (1338243), 2024.

\bibitem[Liang et~al.(2021)Liang, Zou, Liang, Wu, and Yan]{liang2021emotional}
Jing Liang, Yu-Qing Zou, Si-Yi Liang, Yu-Wei Wu, and Wen-Jing Yan.
\newblock Emotional gaze: The effects of gaze direction on the perception of
  facial emotions.
\newblock \emph{Frontiers in psychology}, 12:\penalty0 684357, 2021.

\bibitem[Lu et~al.(2015)Lu, Zheng, Li, and Lu]{lu2015combining}
Yifei Lu, Wei-Long Zheng, Binbin Li, and Bao-Liang Lu.
\newblock Combining eye movements and eeg to enhance emotion recognition.
\newblock In \emph{IJCAI}, pages 1170--1176. Buenos Aires, 2015.

\bibitem[Mehrabian(1996)]{mehrabian1996pleasure}
Albert Mehrabian.
\newblock Pleasure-arousal-dominance: A general framework for describing and
  measuring individual differences in temperament.
\newblock \emph{Current Psychology}, 14:\penalty0 261--292, 1996.

\bibitem[Mohammadi and Vuilleumier(2022)]{AMultiMohammadi2022}
Gelareh Mohammadi and Patrik Vuilleumier.
\newblock A multi-componential approach to emotion recognition and the effect
  of personality.
\newblock \emph{IEEE Transactions on Affective Computing}, 13\penalty0
  (3):\penalty0 1127–1139, 2022.

\bibitem[Park and Lee(2024)]{park2024}
Sungjoon Park and Jihoon Lee.
\newblock Modality effects on emotion perception in english by chinese l2
  english users: An eye-tracking study.
\newblock In \emph{Proceedings of the North American Chapter of the Association
  for Computational Linguistics (NAACL)}, 2024.

\bibitem[Pedregosa et~al.(2011)Pedregosa, Varoquaux, Gramfort, Michel, Thirion,
  Grisel, Blondel, Prettenhofer, Weiss, Dubourg, et~al.]{Scikit-learn}
Fabian Pedregosa, Ga{\"e}l Varoquaux, Alexandre Gramfort, Vincent Michel,
  Bertrand Thirion, Olivier Grisel, Mathieu Blondel, Peter Prettenhofer, Ron
  Weiss, Vincent Dubourg, et~al.
\newblock \emph{Scikit-learn: Machine learning in Python}, 2011.

\bibitem[Rauthmann et~al.(2012)Rauthmann, Seubert, Sachse, and
  Furtner]{rauthmann2012eyes}
John~F Rauthmann, Christian~T Seubert, Pierre Sachse, and Marco~R Furtner.
\newblock Eyes as windows to the soul: Gazing behavior is related to
  personality.
\newblock \emph{Journal of Research in Personality}, 46\penalty0 (2):\penalty0
  147--156, 2012.

\bibitem[Russell(1980)]{Russell1980}
James~A. Russell.
\newblock A circumplex model of affect.
\newblock \emph{Journal of Personality and Social Psychology}, 39\penalty0
  (6):\penalty0 1161--1178, 1980.

\bibitem[Schurgin et~al.(2014)Schurgin, Nelson, Iida, Ohira, Chiao, and
  Franconeri]{schurgin2014eye}
MW Schurgin, J Nelson, S Iida, H Ohira, JY Chiao, and SL Franconeri.
\newblock Eye movements during emotion recognition in faces.
\newblock \emph{Journal of vision}, 14\penalty0 (13):\penalty0 14--14, 2014.

\bibitem[Seikavandi et~al.(2025)Seikavandi, Barrett, and
  Burelli]{seikavandi2025modeling}
Meisam~Jamshidi Seikavandi, Maria~Jung Barrett, and Paolo Burelli.
\newblock Modeling face emotion perception from naturalistic face viewing:
  Insights from fixational events and gaze strategies.
\newblock In \emph{Recent Advances in Deep Learning Applications: New
  Techniques and Practical Examples}. Taylor \& Francis, 2025.

\bibitem[Siegert et~al.(2011)Siegert, B{\"o}ck, Vlasenko, Philippou-H{\"u}bner,
  and Wendemuth]{siegert2011appropriate}
Ingo Siegert, Ronald B{\"o}ck, Bogdan Vlasenko, David Philippou-H{\"u}bner, and
  Andreas Wendemuth.
\newblock Appropriate emotional labelling of non-acted speech using basic
  emotions, geneva emotion wheel and self assessment manikins.
\newblock In \emph{2011 IEEE International Conference on Multimedia and Expo},
  pages 1--6. IEEE, 2011.

\bibitem[Skaramagkas et~al.(2023)Skaramagkas, Giannakakis, Ktistakis, Manousos,
  Karatzanis, Tachos, Tripoliti, Marias, Fotiadis, and
  Tsiknakis]{skaramagkas_review_2023}
Vasileios Skaramagkas, Giorgos Giannakakis, Emmanouil Ktistakis, Dimitris
  Manousos, Ioannis Karatzanis, Nikolaos Tachos, Evanthia Tripoliti, Kostas
  Marias, Dimitrios~I. Fotiadis, and Manolis Tsiknakis.
\newblock Review of {Eye} {Tracking} {Metrics} {Involved} in {Emotional} and
  {Cognitive} {Processes}.
\newblock \emph{Ieee Reviews in Biomedical Engineering}, 16:\penalty0 260--277,
  2023.

\bibitem[Tarnowski et~al.(2020)Tarnowski, Ko{\l}odziej, Majkowski, and
  Rak]{tarnowski2020eye}
Pawe{\l} Tarnowski, Marcin Ko{\l}odziej, Andrzej Majkowski, and Remigiusz~Jan
  Rak.
\newblock Eye-tracking analysis for emotion recognition.
\newblock \emph{Computational intelligence and neuroscience}, 2020, 2020.

\bibitem[Vacaru et~al.(2025)Vacaru, Waters, and Hunnius]{vacaru2025attachment}
Stefania~Victorita Vacaru, Theodore~EA Waters, and Sabine Hunnius.
\newblock Attachment is in the eye of the beholder: a pupillometry study on
  emotion processing.
\newblock \emph{Scientific reports}, 15\penalty0 (1):\penalty0 8015, 2025.

\bibitem[Van~Heijst et~al.(2025)Van~Heijst, Kret, and Ploeger]{van2025basic}
Karlijn Van~Heijst, Mariska~E Kret, and Annemie Ploeger.
\newblock Basic emotions or constructed emotions: Insights from taking an
  evolutionary perspective.
\newblock \emph{Perspectives on Psychological Science}, 20\penalty0
  (3):\penalty0 377--391, 2025.

\bibitem[Wang and Gao(2023)]{wang2023}
Yu Wang and Shiyu Gao.
\newblock Emotion recognition in adaptive virtual reality settings: Challenges
  and opportunities.
\newblock \emph{IEEE Transactions on Affective Computing}, 2023.

\bibitem[Zautra et~al.(2005)Zautra, Affleck, Tennen, Reich, and
  Davis]{zautra2005dynamic}
Alex~J Zautra, Glenn~G Affleck, Howard Tennen, John~W Reich, and Mary~C Davis.
\newblock Dynamic approaches to emotions and stress in everyday life: Bolger
  and zuckerman reloaded with positive as well as negative affects.
\newblock \emph{Journal of personality}, 73\penalty0 (6):\penalty0 1511--1538,
  2005.

\bibitem[Zhang and Wang(2023)]{zhang2023}
Liang Zhang and Minghua Wang.
\newblock Survey of deep emotion recognition in dynamic data using facial,
  speech, and textual cues.
\newblock \emph{Frontiers in Psychology}, 14:\penalty0 10978716, 2023.

\end{thebibliography}
}


\end{document}